\documentclass[final,authoryear,3p,times,twocolumn]{elsarticle}

\usepackage{graphics}

\usepackage{amssymb}
\usepackage{amsthm}
\usepackage{savesym}
\usepackage{amsmath}
\savesymbol{iinit}
\usepackage{wasysym}

\usepackage{lscape}

\journal{Geochimica et Cosmochimica Acta}

\begin{document}

\begin{frontmatter}


\title{{\it Ab initio} prediction of equilibrium boron isotope fractionation between minerals and aqueous fluids at high $P$ and $T$}



\author{
Piotr M. Kowalski$^{1,2}$\corref{cor1}} 
\cortext[cor1]{Corresponding author: Piotr Kowalski, Tel.: +49 2461 61 9356, E-mail: p.kowalski@fz-juelich.de}
\author{
Bernd Wunder$^{1}$ and Sandro Jahn$^{1}$}
\address{
$^1$GFZ German Research Centre for Geosciences, Telegrafenberg, 14473 Potsdam, Germany}
\address{
$^2$Forschungszentrum J\"ulich, Institute of Energy and Climate Research (IEK-6), Wilhelm-Johnen-Strasse, 52425 J\"ulich, Germany}

\begin{abstract}
Over the last decade experimental studies have shown a large B isotope fractionation between
materials carrying boron incorporated in trigonally and tetrahedrally coordinated sites, but the mechanisms 
responsible for producing the observed isotopic signatures are poorly known. 
In order to understand the boron isotope fractionation processes and to obtain a better interpretation of the 
experimental data and isotopic signatures observed in natural samples,
we use first principles calculations based on density functional theory in conjunction with {\it ab initio} molecular dynamics
and a new pseudofrequency analysis method
to investigate the B isotope fractionation
between B-bearing minerals (such as tourmaline and micas) and aqueous fluids containing $\rm H_3BO_3$ and $\rm H_4BO_4^-$ species.
We confirm the experimental finding that the isotope fractionation is mainly driven by the coordination of the fractionating 
boron atoms and have found in addition that the strength of the produced isotopic signature is strongly correlated with the B-O bond length.
We also demonstrate the ability of our computational scheme to predict the isotopic signatures of fluids at extreme pressures
by showing the consistency of computed pressure-dependent $\beta$ factors 
with the measured pressure shifts of the B-O vibrational frequencies
of $\rm H_3BO_3$ and $\rm H_4BO_4^-$ in aqueous fluid.
The comparison of the predicted with measured fractionation factors between boromuscovite and neutral fluid
confirms the existence of the admixture of tetrahedral boron species in neutral fluid at high $P$ and $T$ found experimentally,
which also explains the inconsistency between the various measurements on the tourmaline-mica system reported in the literature.
Our investigation shows that 
the calculated equilibrium isotope fractionation factors have an accuracy comparable to the experiments and give unique and valuable
insight into the processes governing the isotope fractionation mechanisms on the atomic scale. 

\end{abstract}


\end{frontmatter}


\section{Introduction}
\label{}

The use of isotopes as geochemical tracers depends upon the existence of
reliable $P-T$-dependent equilibrium isotope fractionation data between solids, fluids and
melts. The common method used for determination of such data is to conduct
experiments, where the phases of interest are equilibrated at a range of
different conditions and individually measured for their equilibrium isotopic
compositions. Recently, with the development of computational methods, software and 
increase in hardware performance it is also possible to simulate and 
compute the isotope fractionation with {\it ab initio} methods.
These computational considerations complement the experimental effort
and provide information on the mechanisms governing the equilibrium isotope 
fractionation processes on the atomic scale.
By establishing an efficient computational approach for materials at high $P$ and $T$
and testing its reliability by computing Li isotope fractionation between minerals 
and aqueous fluids, \citet{KJ11} have shown that {\it ab initio} methods can provide
a reliable estimation of equilibrium isotope fractionation factors at an accuracy level
comparable to experiments.
Motivated by the encouraging results for the fractionation of lithium isotopes we applied 
our new method to study boron isotope fractionation. The main goals
of the present study are to make theoretical predictions and obtain a better understanding of the B isotope fractionation process between
tourmaline, B-bearing mica and fluids at various pressures and temperatures,
to compare these data to results from recent {\it in situ} experimental studies \citep{WM05,MW08} 
and measurements of isotopic signatures in natural samples \citep{KM11,M05,H02}
and to investigate the underlying mechanisms driving the boron isotope fractionation processes
between the considered materials.

With two stable isotopes $^{10}$B and $^{11}$B, of relatively large mass
difference of about $10\,\%$, boron isotopes strongly fractionate during geological
processes, thus leading to natural $\rm\delta^{11}B$-variations ranging 
from $-30$ to $\rm+60\,\permil$ \citep{B93}. Therefore, B isotopes are ideal 
for the distinction of
different geological environments and for quantifying mass transfer
processes, e.g. in the range of subduction zones. First-order criteria
driving isotope fractionation in Earth materials are differences in
coordination and in the bonding environments of coexisting phases. The
lighter isotope usually preferentially occupies the higher coordinated site,
which is generally accompanied with a longer cation-anion bond length and
weaker bond strength \citep{SMH09,WJ11}.

Tourmaline has an extensive chemical variability and is the most widespread
borosilicate in various rocks over a large range of bulk compositions. It
has a large $P-T$-stability which ranges from surface conditions to high
pressures and temperatures of at least 7.0 GPa and about $\rm 1000^{o}C$ as
determined experimentally for dravitic tourmaline \citep{K95}. Most
tourmaline minerals contain 3 boron atoms per formula unit (pfu) with B in trigonal planar
coordination (B$^{[3]}$). High pressure combined with an Al-rich environment can
lead to the formation of olenitic tourmaline with significant amounts of
excess B substituting for Si at the tetrahedral site (B$\rm^{[4]}$). The highest amounts
of up to 1.2 B$^{[4]}$ pfu have been found in olenitic tourmaline from Koralpe,
Austria \citep{EP97}. Al-rich tourmaline with up to 2.2 B$^{[4]}$ pfu was
synthesized experimentally at $\rm 2.5\,GPa$, 600$^{\rm o}$C \citep{SW00}. The
maximum possible amount of B$\rm ^{[4]}$ in olenitic tourmaline is limited 
to three B$\rm^{[4]}$ pfu, due to structural and crystal chemical reasons.

As tourmaline is not stable at basic pH \citep{ML89},
the dominant B-species of fluids coexisting with tourmaline is $\rm B(OH)_3$.
Therefore, due to the absence of change in boron coordination, B-isotopic
fractionation between B$\rm^{[4]}$-free tourmaline and fluid ($\rm \Delta^{11}B$(tour-fl)) should
be small. However, at $0.2\rm\,GPa$, the experimentally determined $\rm \Delta^{11}B$(tour-fl)-values of 
$-2.5\pm0.4\,\permil$ at 400$\rm^{o}C$ and $-0.4 \pm 0.4\,\permil$ at 700$\rm^{o}C$ \citep{MW08} 
suggest small but significant differences in the B-O(H) bond strength between
tourmaline and the neutral fluids. Incorporation of $\rm B^{[4]}$ into tourmaline is
expected to significantly increase the fractionation of boron isotopes between 
the mineral and aqueous fluid (i.e. increase of $|\rm \Delta^{11}B|$(tour-fl)). In this contribution we
present calculated $P-T$-dependent data on B-isotope fractionation between
B$\rm ^{[4]}$-bearing tourmaline and fluid, which so far are not available from
experiments.

Despite of its low boron content of up to maximum values of $\rm 270\,ppm$
\citep{DH93}, potassic white mica is probably the main host of
boron in metasedimentary and metabasaltic blueschists and eclogites, because
of its high modal abundance and B-incompatibility in all other stable
minerals of these rocks \citep{BRS98}. During subduction
the modal amount of mica continuously decreases by dehydration reactions and
the chemistry of residual micas is shifted towards phengitic compositions.
Phengitic mica has an extended stability, extending as deep as $300\rm\,km$ within cold
subduction zones \citep{S96}. In contrast to most tourmalines, boron
is tetrahedrally coordinated in mica, where it substitutes for aluminum.
Due to the coordination change from mostly three-fold coordinated in
near-neutral fluids 
\citep{STH05} 
to B$\rm^{[4]}$ in mica, B-isotopic fractionation between B-bearing mica and
such fluids is much larger than for tourmaline -- fluid.
$\rm \Delta^{11}B$(mica-fl)-values determined experimentally at $P=3.0\rm\, GPa$,
are $\rm-10.9\pm1.3\,\permil$ at 500$\rm^{o}C$ and $\rm-7.1\pm0.5\,\permil$
at 700$\rm ^{o}C$ \citep{WM05}. Such a strong B-isotope fractionation and its
pronounced $T$-dependence, in combination with the continuous dehydration of micas
during ongoing subduction and boron transport via fluids into mantle wedge
regions of arc magma-formation, probably determines the boron variations and
B-isotopic signatures in volcanic arcs \citep{WM05}.

Using {\it in situ} Raman spectroscopic
measurements of near-neutral B-bearing fluids, \citet{STH05} observed
a significant amount of 4-fold B-species at high $P$ and $T$. The abundance of these species 
increases with temperature and pressure and at $T\rm=800\,K$ and $P\rm=1.9\,GPa$ there should be 
a considerable amount ($\rm 15-30\%$) of $\rm B^{[4]}$ species in the fluid.
Such a significant amount of tetrahedrally coordinated B-species in high temperature and pressure
fluids should affect solid--fluid B-isotopic fractionation, which 
we investigate in our calculations. In the light of this, we also discuss
recently determined B-isotope data from coexisting natural tourmaline and
B-bearing mica \citep{KM11,M05,H02}, which show slight inconsistency with {\it in situ} measurements of \citet{WM05} and \citet{MW08}.

Reliable {\it ab initio} computational methods to predict isotope fractionation factors have been established recently.
Several groups have proved that such calculations can contribute towards understanding geochemical mechanisms responsible for production 
of isotope signatures \citep{D97,YMMW01,S04,DK08,HS08,ML09,ML07,SMH09,Z09,HS10,RC10,RB10,Z09,Z10,KJ11}.
The majority of these works, however, concentrate on the computation of the stable isotope fractionation between
various, mostly simple crystalline minerals, and the aqueous solutions are usually computed using 
an isolated cluster containing fractionating species and a hydration shell (e.g. \citet{Z10,Z09,HS10,RC10,DK08,HS08,RB10,Z05}). However, 
aqueous solutions at high pressure and temperature must be computed with caution as discussed in \citet{KJ11}.
This is because the distribution of cation coordination and cation-oxygen bond lengths that affect the isotope fractionation \citep{BM47}
is driven by the dynamics of the system and change under compression \citep{JW09,WJ11,KJ11}.
The only recent {\it ab initio}
work, besides \citet{KJ11}, that accounts for the dynamical effects on the isotope fractionation in fluid is by \citet{RB07}
who considered boron equilibrium isotope fractionation between $\rm B(OH)_3$ and $\rm B(OH)_4^-$ species in aqueous solution.
They performed {\it ab initio} molecular dynamics simulations of these fluids and attempted to use the vibrational density of states,
derived through the Fourier transform of the velocity auto-correlation function, as an input for the calculation of the 
$^{11}$B/$^{10}$B isotope fractionation coefficient. However, the resulting fractionation factor $\alpha=0.86$  
happened to be much lower than the experimental value of
$\alpha=1.028$. Interestingly, the discrepancy between experiment and theory is cured by quenching the selected configurations
along the molecular dynamics trajectory and 
computing the harmonic frequencies. The fractionation factor derived using these frequencies
exactly reproduces the experimental value.

In our approach both solids and fluids are treated as extended systems by application of periodic boundary conditions in all three spatial directions,
which is crucial to model high pressure materials. Large enough supercells are chosen to avoid significant interaction between atoms and their periodic images,
as well as to reduce the number of k-points and q-vectors for sampling the Brillouin
zones and the phonon spectra of the crystals, respectively (for a liquid or fluid, both the Brillouin zone and phonons are not
defined).
In our investigation we use cells of at least $7\rm\,\AA$ width in each spatial dimension.
A representative statistical sampling of the fluid structure is obtained by performing
Car-Parrinello molecular dynamics simulations \citep{CPMD1}.
For the calculation of the isotope fractionation factors in fluids, several random snapshots from the simulation runs are chosen.
The force constants acting on the fractionating element and the resulting fractionation factors are then obtained for each ionic configuration, and
the relevant fractionation factor for the boron species in the fluid is computed as an average over the whole set of considered geometries.
In \citet{KJ11} it was shown that in line with the \citet{BM47} approximation, considering the force constants acting on the fractionating atom only
leads to a satisfactory estimation of the Li isotope fractionation factors for high temperature fluids and minerals.
Here, we will show that approximating the vibrational spectrum by the three pseudofrequencies derived from the force constants 
allows for further improvement of the accuracy of the predicted isotope fractionation factors, especially at lower temperatures.

In this contribution we present the theoretical prediction of B isotope fractionation factors between B bearing aqueous fluids and solids, specifically 
tourmaline and B-muscovite, for which experimental data and measurements on natural samples are available for comparison \citep{WM05,MW08,KM11}.
We will show that the application of {\it ab initio} methods to B-bearing crystalline solids and fluids not only provides unique insight into the
mechanisms driving equilibrium B-isotope fractionation on the atomic scale, but helps in proper interpretation of the data.

\begin{table*}[t]
\caption{The computed and measured vibrational frequencies for 
$\rm H_3BO_3$ and $\rm H_4BO_4^-$ molecules reported for two different boron isotopes
B$^{11}$/B$^{10}$. Only the frequencies affected by the isotopic substitution
are reported. The units are $\rm cm^{-1}$.}
\label{T1a}      
\centering          
\begin{tabular}{lcccccc}     
\hline\hline       
method & $\rm H_3BO_3$ & & $\rm H_4BO_4^-$ & & & \\
\hline                    
theoretical:  &  &  &   &  &  &   \\
BLYP (this work) & 656/680 & 1366/1408  & 793/807  & 870/886 & 1051/1063 & 1132/1154   \\
BLYP $^1$ & 642/667 & 1390/1437  & 807/821  & 916/947 & 1020/1025 & 1156/1179   \\
HF/6-31G$^*$ $^2$ & 736/764 & 1562/1615  &   &  &  &   \\
HF/aug-cc-pVDZ$^*$ $^3$ & \hspace{4mm}-/754 & \hspace{6mm}-/1531  &   &  &  &   \\
BP86/aug-cc-pVDZ$^*$ $^3$ & \hspace{4mm}-/655 & \hspace{6mm}-/1412  &   &  &  &   \\
B3LYP/aug-cc-pVDZ$^*$ $^3$ & \hspace{4mm}-/684 & \hspace{6mm}-/1447  &   &  &  &   \\
MP2/aug-cc-pVDZ$^*$ $^3$ & \hspace{4mm}-/684 & \hspace{6mm}-/1435  &   &  &  &   \\
CCSD(T)/aug-cc-pVDZ$^*$ $^3$ & \hspace{4mm}-/684 & \hspace{6mm}-/1438  &   &  &  &   \\
experimental: &  &  &   &  &  &   \\
solution $^4$ & 632/666 & 1412/1454  &   & 937/975 &  &   \\
vapour $^5$ & 674/700 & 1429/1477  &   &  &  &   \\    
solution $^6$ & 639/668 & 1428/1490  &   & 947/-\hspace{4mm} &  &   \\ 
vapour $^7$ & 666/692 & 1415/1472  &   &  &  &   \\  
vapour $^8$ & 675/701 & 1426/1478  &   &  &  &   \\  
solution and vapour $^9$ & 639-675/- & 1421-1450/- & & 935-958/- & & \\
\hline                  
\end{tabular}
\\
References: $^1$\citet{Z05}, $^2$\cite{LT05}, $^3$\cite{RB10}, $^4$\cite{SV05}, 
$^5$\cite{G91}, $^6$\cite{O00,LT05}, $^7$\cite{AB92}, $^8$\cite{OY88}, $^9$\cite{Z05} and references hereafter.
\end{table*}

\begin{table*}[t]
\caption{$1000(\beta-1)$ factors for isolated $\rm H_3BO_3$ and $\rm H_4BO_4^-$ molecules at $T\rm=300\,K$ obtained using three
methods: (1) the full frequency spectrum and Equation \ref{beta}, (2) {\it the single atom approximation} of \citet{KJ11}, 
(3) {\it the single atom approximation} with {\it the pseudofrequencies} and equation \ref{beta}. The units are $\permil$. The $\Delta\beta/\beta_{\rm BM}$
is the check of condition given by Eq. \ref{TESTP} and indicates the improvement of the method (3) over method (2) expressed in $\%$.}
\label{T1}      
\centering          
\begin{tabular}{lrcccrccc}     
\hline\hline       
T (K) &  meth. (1) & meth. (2) & meth. (3) & $\Delta\beta/\beta_{BM}\,[\%]$ & meth. (1) & meth. (2) & meth. (3) & $\Delta\beta/\beta_{BM}\,[\%]$\\
\hline                    
     & $\rm H_3BO_3$ & & & & $\rm H_4BO_4^-$ & & & \\
 300 &  211.3  & 283.1 &  225.1  & 20.5 & 167.3 &  203.4  & 175.2 & 13.9  \\
 600 &  64.1  & 70.8  &  65.6 & 7.3 & 48.1 &  50.9  & 48.6 & 4.5  \\
 800 &  37.6 & 39.8  &  38.1 & 4.3 & 27.8 &  28.6  & 27.9 & 2.5  \\
1000 &  24.6  & 25.4  &  24.7 & 2.8 & 18.0 &  18.3  & 18.0 & 1.7 \\
     & Dravite & & & & Boromuscovite 1M & & & \\
 300 &  202.8  & 267.7 & 216.0 & 19.3 & 139.1 &  160.8  & 142.2 & 11.6  \\
 600 &   60.7  &  66.9 &  62.4 & 6.7 & 38.6 &  40.2  & 38.8 & 3.5  \\
 800 &   35.5  &  37.6 &  36.1 & 4.0 & 22.1 &  22.6  & 22.2 & 1.8  \\
1000 &   23.2  &  24.1 &  23.5 & 2.5 & 14.3 &  14.5  & 14.3 & 1.4  \\
\hline                  
\end{tabular}
\\
\end{table*}

\section{Computational approach}

\subsection{Theoretical model \label{TM}}
\subsubsection{The single atom approximation: \citet{BM47} approach}
Mass-dependent equilibrium isotope fractionation is driven by the change in molecular and crystalline vibration frequencies
resulting from the different masses of the isotopes.
The fractionation between species and an ideal monoatomic gas is called the $\beta$ factor. In the harmonic approximation it is given by the formula
\citep{BM47,U47,CCH01}:
\begin{equation} \beta=\prod_{i=1}^{N_{dof}}\frac{u_{i}^{*}}{u_{i}}\exp{\left[ \frac{(u_{i}-u_{i}^*)}{2}\right]}\frac{1-\exp(-u_i)}{1-\exp(-u_i^*)}, \label{beta}\end{equation}
where $u={h\nu_i}/{k_BT}$, $h$ is the Planck constant, $\nu_i$ 
is the vibrational frequency of the $i$-th degree of freedom,
$k_B$ is the Boltzmann constant,
$N_{dof}$ is the number of degrees of freedom, which for $N$ being the number of atoms in 
the considered system (molecule, mineral or fluid)
is equal to $3N-5$ for a diatomic molecule, $3N-6$ for multiatomic molecules and $3N$ for crystals, 
and a star symbol marks the heavier isotope.
The fractionation factor between two substances A and B, $\alpha_{A-B}$ 
is computed as the ratio of the relevant $\beta$ factors, which
is well approximated by the differences in the $\beta$ factors:
\begin{equation} \alpha_{A-B}=\beta_A/\beta_B,\,\Delta_{A-B}\cong 1000\ln\beta_A-1000\ln\beta_B [\permil].\end{equation}
The calculation of the $\beta$ factor requires only knowledge of the vibrational properties 
of the considered system computed for the two different isotopes.
However, computation of the whole vibrational spectra of complex, multiparticle 
minerals or fluids requires substantial computational resources and is currently limited to systems containing a few dozens of atoms or less.
In our recent work \citep{KJ11} we proposed to use an efficient method for computing the high temperature isotope fractionation factors between complex materials such as fluids and crystalline solids,
which requires the knowledge of the force constants acting upon the fractionating element only. 
The $\beta$ factor (Eq. \ref{beta}) can be then approximated by \citep{BM47,KJ11}:
\begin{equation} 
\beta\simeq 1+\sum_{i=1}^{N_{dof}}\frac{u_i^2-u_i^{*2}}{24}=1+\frac{\Delta m}{m m^*}\frac{\hbar^2}{24 k_B^2T^2}\sum_{i=1}^3 A_i, \label{BAPX} 
\end{equation}
where $A_i$ are the force constants acting on the isotopic atom in the three perpendicular spatial directions (x, y and z),
$\Delta m=m^*-m$, where $m$ and $m^*$ are the masses of the lighter and heavier isotopes of the fractionating element.
As the computation of the $\beta$ factors from formula \ref{BAPX} requires the knowledge of properties of the fractionating element only
we will call such an approach {\it the single atom approximation} throughout the paper.
The validity criteria restricts the usage of the formula to frequencies $ \nu\,{\rm[cm^{-1}]}\lesssim1.39\,T\rm{[K]}$ (assuming $u<2$, see Fig. 1 of \citet{BM47}). 
We are interested in temperature range $800$-$1000\rm\,K$. The highest vibrational frequency of the modes involving movement of B atoms for $\rm H_3BO_3$ is $\sim1400\,\rm cm^{-1}$
and of $\rm H_4BO_4^{-1}$ is $<1200\,\rm cm^{-1}$ (Table \ref{T1a}). In case of $\rm H_3BO_3$, the single atom approximation may produce an error of $2.2\, \permil$ in the $\beta$ factor at $T\rm=800\,K$.
The relevant error for $\rm H_4BO_4^{-1}$ is $0.8\, \permil$ (Table \ref{T1}). A further improvement to the method is therefore desired.

\subsubsection{The single atom approximation with pseudofrequencies: our improvement \label{SAA}}
We will show that the error of {\it the single atom approximation} can be substantially reduced if one uses the three frequencies $\bar{\nu_i}$ derived from the force constants acting on the
fractionating element ($\bar{\nu}_i^2=A_i/4\pi^2m$). We call them {\it``pseudofrequencies"}, and compute the $\beta$ factors using formula \ref{beta}.
In the following we present the formal justification of such an approach. According to \citet{BM47}, equation \ref{beta} for small $\Delta u_i=u_i-u_i^*$ reduces to (\citet{BM47}, Eq. 11a):
\begin{equation} \beta=1+\sum_{i=1}^{N_{dof}}\left (\frac{1}{2}-\frac{1}{u_i}+\frac{1}{\exp(u_i)-1}\right )\Delta u_i. \label{BEXP}\end{equation}
The Taylor expansion of the function appearing under the summation sign is:
\begin{equation*} G(u)=\frac{1}{2}-\frac{1}{u}+\frac{1}{\exp(u_i)-1}        \end{equation*}
\begin{equation}=\frac{u}{12}-\frac{u^3}{720}+\frac{u^5}{30240}-\frac{u^7}{1209600}+... .\end{equation}
When we consider just the first term of the expansion the $\beta$ factor is:
\begin{equation*} \beta=1+\sum_i^{N_{dof}} \frac{u_i}{12}\Delta u_i \end{equation*}
\begin{equation}=1+\sum_{i=1}^{N_{dof}} \frac{\Delta u_i^2}{24}=1+\sum_{i=1}^{N_{dof}} \frac{u_i^2-u_i^{*2}}{24},\end{equation}
which is exactly equation \ref{BAPX}.

\begin{table}[t]
\caption{The three pseudofrequencies of the B atom derived 
for two selected molecules and two crystalline solids. The units are $\rm cm^{-1}$.}
\label{T1b}      
\centering          
\begin{tabular}{lccc}     
\hline\hline       
species & & &  \\
\hline                    
$\rm H_3BO_3$ & 644/675 & 1129/1184  & 1130/1185     \\
dravite & 618/648 & 1085/1138  & 1120/1174     \\
$\rm H_4BO_4^-$ & 809/848 & 845/886  & 872/915     \\
boromuscovite 1M & 722/757 & 723/758  & 801/840     \\
\hline                  
\end{tabular}
\end{table}

\begin{figure*}[t]
\includegraphics[angle=270,width=3.5in]{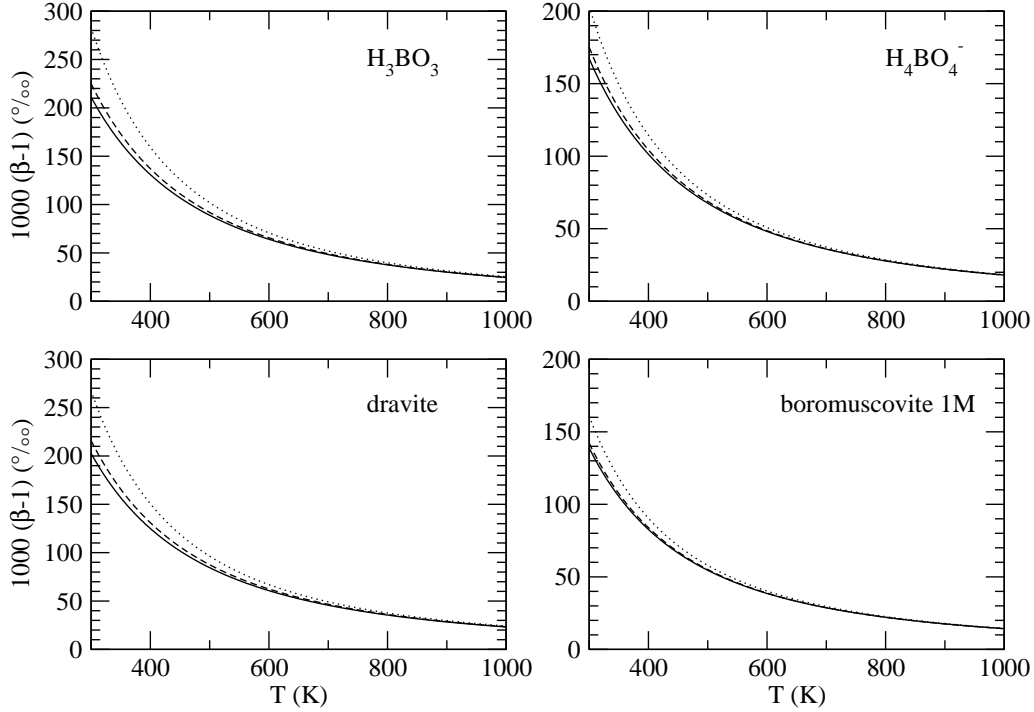}
\caption{ $\beta$ factors for isolated $\rm H_3BO_3$ and $\rm H_4BO_4^-$ molecules as well as dravite and boromuscovite 1M 
crystalline solids.
The lines represent the results obtained
using: (1) the full frequency spectrum and Eq. \ref{beta} (solid), (2) {\it the single atom approximation } 
of \citet{BM47,KJ11}, Eq. \ref{BAPX} (dotted) and (3) the method described in the text with {\it the pseudofrequencies} and Eq. \ref{beta} (dashed lines). \label{F1}}
\end{figure*}

\begin{table*}[t]
\caption{The lattice parameters of the investigated B-bearing crystalline solids. 
$N_{atoms}$ is the number of atoms in the modeled supercell.}
\label{T2}      
\centering          
\begin{tabular}{lcccccc}     
\hline\hline       
& boromuscovite 1M$^1$ & boromuscovite 2M1$^1$ & dravite$^2$ & olenite$^3$ \\
\hline                    
 a (\AA) &  10.204 & 10.180  & 15.945  & 15.5996   \\
 b (\AA) &  8.788 &  8.822  & 15.945  & 15.5996   \\
 c (\AA) & 10.076 & 19.8189  & 7.210  &  7.0224   \\
 $\alpha$ ($^\circ$) & 90  & 90   & 90  & 90  \\
 $\beta$  ($^\circ$) & 101.23  & 95.62  & 90  & 90  \\
 $\gamma$ ($^\circ$) & 90  & 90   & 90  & 120  \\
 $N_{atoms}$ & 84 & 168& 163 & 162\\
\hline                  
\end{tabular}
\\
References: $^1$\cite{LH95}, $^2$\cite{EM10}, $^3$\cite{MB02}
\end{table*}

Let us consider the Taylor expansion of the different estimations of $\beta$ factors. Equation \ref{BEXP} then reads:
\begin{equation} \beta=\beta_{exact}=1+\sum_{i=1}^{N_{dof}}\left (\frac{u_i}{12}-\frac{u_i^3}{720}+... \right )\Delta u_i \label{B11}.\end{equation}
The \citet{BM47} approximation given by equation \ref{BAPX} reads:
\begin{equation} \beta\sim\beta_{BM}=1+\sum_{i=1}^{N_{dof}}\frac{u_i}{12}\Delta u_i \label{B111},\end{equation}
and the proposed approximation based on pseudofrequencies is:
\begin{equation} \beta\sim\beta_{pseudo}=1+\sum_{i=1}^3\left (\frac{\bar{u}_i}{12}-\frac{\bar{u}_i^3}{720}+... \right )\Delta \bar{u}_i. \label{B22}\end{equation}
In the above equation $\bar{u_i}=h\bar{\nu_i}/k_BT$.
Next we check how this approximation compares to the \citet{BM47} approximation given by equations \ref{BAPX} and \ref{B111}. In order to make a comparison we derive
the differences between the two approximate expressions $\beta_{BM}$, $\beta_{pseudo}$ and the exact one $\beta_{exact}$ (Eq. \ref{B11}).
In the case of the \citet{BM47} approximation we have:
\begin{equation*} \Delta\beta_{BM}=\beta_{BM}-\beta_{exact}=\left (\sum_i^{N_{dof}}\frac{u_i}{12} -\sum_i^{N_{dof}} \left (\frac{u_i}{12}-\frac{u_i^3}{720}+... \right)\right)\Delta u_i \end{equation*}
\begin{equation} =\sum_i^{N_{dof}}\left ( \frac{u_i^3}{720}-... \right )\Delta u_i\end{equation}
and having from equations \ref{BAPX} and \ref{B111} that 
\begin{equation} \sum_i^{N_{dof}}\frac{u_i}{12}\Delta u_i=\sum_i^{3}\frac{\bar{u_i}}{12}\Delta \bar{u_i} \label {BMBM} \end{equation}
in the case of the proposed approximation we get:
\begin{equation*} \Delta\beta_{pseudo}=\beta_{pseudo}-\beta_{exact}=\sum_{i=1}^3\left (-\frac{\bar{u}_i^3}{720}+...\right )\Delta \bar{u}_i \end{equation*}
\begin{equation*} +\sum_i^{N_{dof}}\left (\frac{u_i^3}{720}-...\right)\Delta u_i=\end{equation*}
\begin{equation} =\sum_{i=1}^3\left (-\frac{\bar{u}_i^3}{720}+...\right )\Delta \bar{u}_i+\Delta\beta_{BM} \label{BPS}\end{equation}
Because relation $G(u)<\frac{u}{12}$ holds for any $u$  (see \citet{BM47}, Fig. 1), the function
\begin{equation} \sum_{i=1}^3\left (-\frac{\bar{u}_i^3}{720}+...\right )\Delta u_i=\sum_{i=1}^3\left ( G(\bar{u_i})-\frac{\bar{u_i}}{12}\right ) \Delta \bar{u_i}<0\end{equation}
and $\Delta\beta_{pseudo}<\Delta\beta_{BM}$. On the other hand the expression for $\Delta\beta_{pseudo}$ is given by a difference of the higher order terms of the Taylor expansions of the 
two expressions for the $\beta$ factor. In the considered cases the values of pseudofrequencies are similar to the real frequencies
that are affected upon B isotope substitution in a given B-bearing system. This can be seen by comparing the pseudofrequencies computed for the selected cases of B-bearing molecules 
and crystalline solids considered here and reported in Table \ref{T1b} with the real frequencies given in Table \ref{T1a}.
This indicates that the two terms of opposite signs in Eq. \ref{BPS} should be similar 
in value and cancel out to a great extent, so $|\Delta\beta_{pseudo}|<<\Delta\beta_{BM}$. Therefore the approach proposed here to compute the $\beta$ factor based on pseudofrequencies and Eq. \ref{beta}
should give a better approximation to the exact $\beta$ factors than equation \ref{BAPX}, which we will show in section \ref{TEST}.
The difference to the \citet{BM47} approximation is given by:
\begin{equation} \Delta\beta=1+\sum_{i=1}^3\frac{\bar{u_i}}{12} \Delta \bar{u_i}-\beta_{pseudo}\end{equation}
and can be easily computed for any considered system. We assume that the pseudofrequency-based approach to the computation of $\beta$ factors is applicable 
if it is just a correction to equation \ref{BAPX}, i.e. when:
\begin{equation} \Delta\beta<<\sum_{i=1}^3\frac{\bar{u_i}}{12} \Delta \bar{u_i}. \label{TESTP} \end{equation}

\begin{figure*}[t]
\includegraphics[angle=270,width=3.5in]{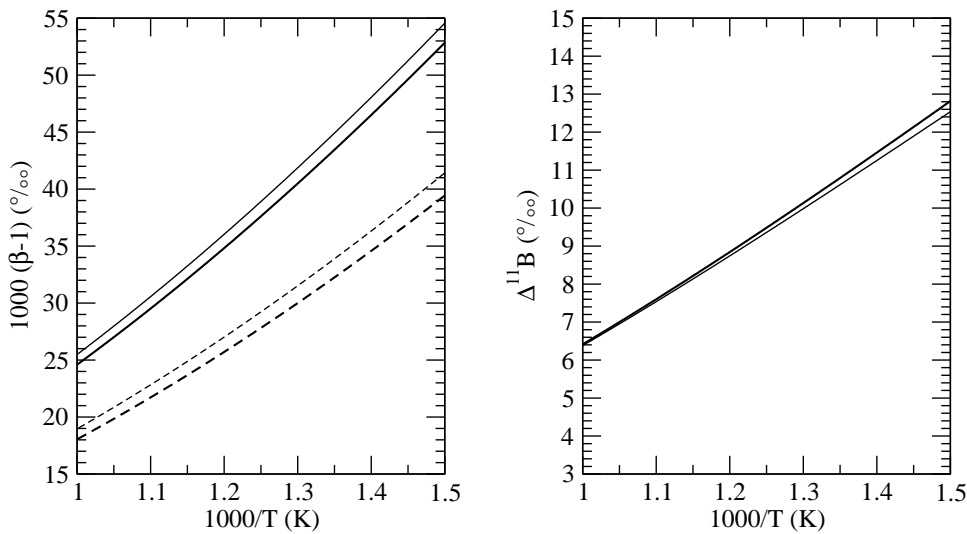}
\caption{Left panel: $\beta$ factors for $\rm H_3BO_3$ (solid line) and $\rm H_4BO_4^-$ (dashed lines) in the gas phase. 
Right panel: fractionation factors between $\rm H_3BO_3$ and $\rm H_4BO_4^-$ in gas phase. The thick lines represent our results,
while the thin lines represent the results obtained using frequencies of \citet{Z05}. \label{F6}}
\end{figure*}

We also note that the proposed approach satisfies the Redlich-Teller product rule \citep{R35} when the mass of considered isotope $m$ 
is much smaller than the mass of the whole considered system $M$, namely,
\begin{equation} \prod_{i=1}^{N_{dof}}\frac{u_i^*}{u_i}=\left(\frac{M^*}{M}\frac{m}{m^*}\right)^{3/2}\approxeq\left(\frac{m}{m^*}\right)^{3/2}= \prod_{i=1}^{3}\frac{\bar{u}_i^*}{\bar{u}_i}.\end{equation}
We notice that we could force the strict conservation of the Redlich-Teller product by just adjusting the ratios of $\bar{u_i}/\bar{u_i^*}$.
However, such a modification would not preserve relation \ref{BMBM} 
and the pseudofrequency approach would not recover exactly the high temperature limit (Eq. \ref{BAPX}) of the exact solution (Eq. \ref{beta}),
which is a more important constraint to fulfill strictly by the proposed approximation.

\subsection{Representation of solids}

In this paper we investigate the boron isotope fractionation between dravite, olenite and boromuscovite minerals and aqueous fluids.
The solids were represented by large cells containing at least 84 atoms. The number of atoms used in the 
crystal calculations together with the lattice parameters of modeled crystals are summarized in Table \ref{T2}. 
The lattice parameters and chemical compositions of the modeled crystalline solids are the experimental values measured at ambient conditions 
found in the literature.
Dravite is the crystalline solid which was used in the experiments on tourmaline by \citet{MW08}. 
The chemical composition of the supercell used in the investigation is $\rm Na_{3} Mg_{9} Al_{18} (Si_{18}O_{54})(B^{[3]}O_{3})_{9}(OH)_{12}$
with structural data of \citet{MB02}.
Olenite can contain B in both trigonal and tetragonal sites. The modeled structure is that of \citet{EM10}. 
The chemical composition of the unit cell used in the investigation is $\rm Na Al_{3} Al_{6} (Si_{4}B_{2}^{[4]}O_{18})(B^{[3]}O_{3})_{3}(OH)_{3}O$.
For boromuscovite, the 1M and 2M1 crystal structures of \citet{LH95} were used.
In the isotope fractionation experiments of \citet{WM05} boromuscovite forms two polytypes, 
1M and 2M1, with relative abundances of $10\%$ and $90\%$ respectively.
In boromuscovite B occupies the 4-fold coordinated site occupied mainly by Si atoms. 
The constructed model constitutes a 2x1x1 supercell of elementary chemical composition $\rm K Al_{2} (B^{[4]} Si_{3} O_{10})(OH)_{2}$.

\subsection{Representation of aqueous solution}

The aqueous solution was represented by a periodically repeated box containing up to 64 water molecules and one $\rm H_3BO_3$ or $\rm H_4BO_4^-$ molecule. 
The pressure and temperature conditions were chosen to be close to the experimental conditions of \citet{WM05} and \citet{MW08}.
The pressure of the aqueous solution for a given temperature and volume was calculated according to the equation of state of
\citet{WP02}. The {\it ab initio} molecular dynamics simulations (AIMD) of aqueous fluids were performed for fixed temperature and volume 
using the Car-Parrinello scheme \citep{CPMD1}. The temperature during each run was controlled by a Nos{\'e}--Hoover chain thermostat \citep{NK83,H85}.
For each $T-V$ conditions at least $10\rm\,ps$ long trajectories were generated with an integration step of $0.12\rm\, fs$.

\begin{figure*}[t]
\includegraphics[angle=270,width=3.5in]{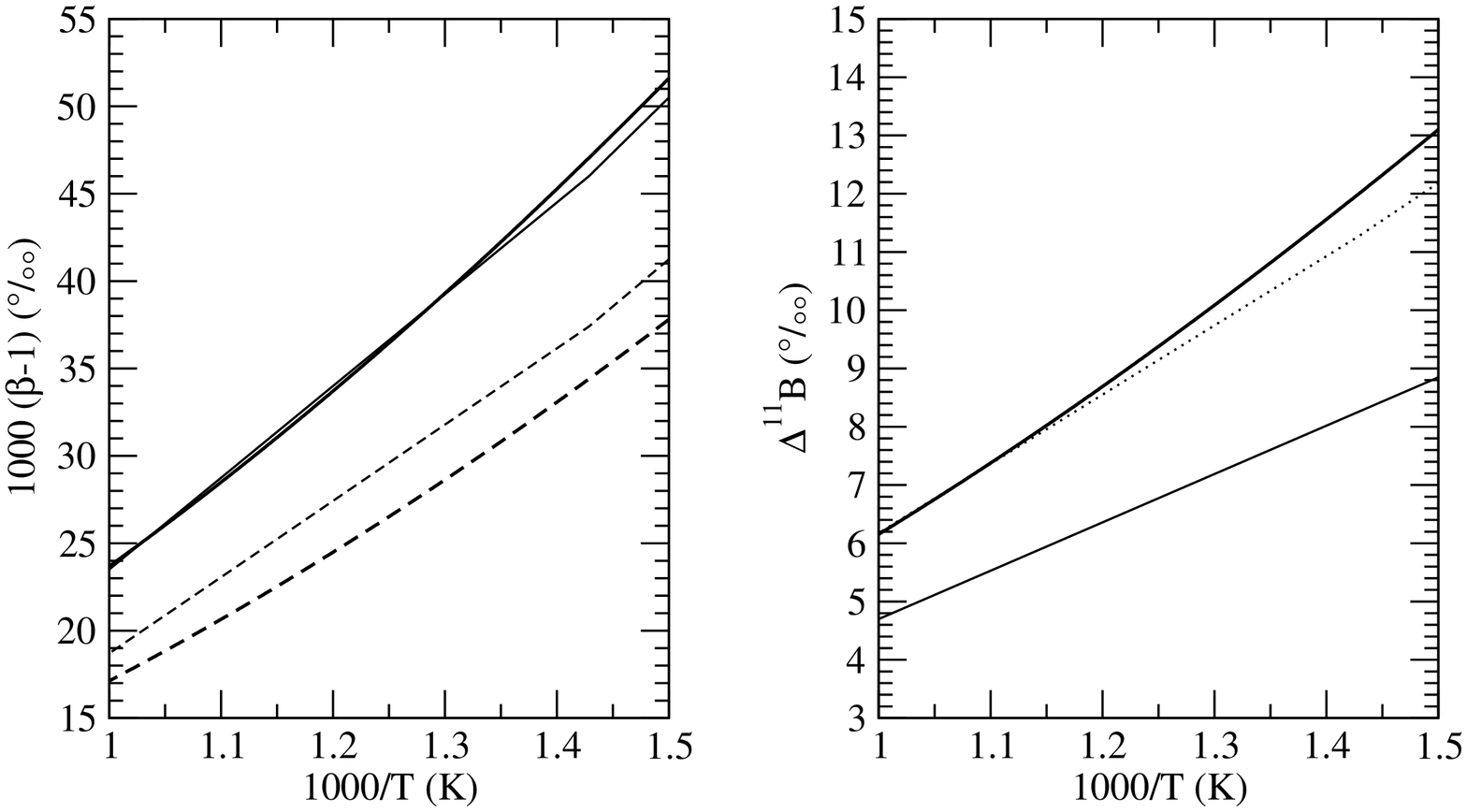}
\caption{Left panel: $\beta$ factors for $\rm H_3BO_3$ (solid line) and $\rm H_4BO_4^-$ (dashed lines) in aqueous solution. 
Right panel: fractionation factors between $\rm H_3BO_3$ and $\rm H_4BO_4^-$ in aqueous solution. The thick lines represent our results,
while the thin lines represent the results of \citet{SV05} obtained using harmonic frequencies (their Table 2).
The dotted line represents the corrected \citet{SV05} results. The correction is made by comparing the work of \citet{RB10} 
and the correction derivation procedure is discussed in the text. \label{F5}}
\end{figure*}

\subsection{Computational technique\label{CA}}

The calculations of pseudofrequencies and
$\beta$ factors 
for solids and aqueous solutions were performed by applying density functional theory (DFT) methods, 
which are currently the most efficient methods allowing for treating extended many particle systems quantum-mechanically. 
We used the planewave DFT code CPMD \citep{CPMD2}, which is especially suited for {\it ab initio} simulations of fluids,
the BLYP exchange-correlation functional \citep{BECKE88,LEE88} and norm-conserving Goedecker pseudopotentials for the description 
of the core electrons \citep{GOE96}. One advantage of using the BLYP functional is that it usually gives harmonic frequencies that most closely 
resemble the observed frequencies of benchmark chemical systems \citep{FS95,AZZ10} \footnote{If the computed harmonic frequencies 
are closer to the observed, anharmonic frequencies than to the real harmonic frequencies of the computed system
the resulted fractionation factor $\beta$ (Eq. \ref{beta}) computed from
these frequencies should be closer to the real fractionation factor than the $\beta$ computed on a set of accurate harmonic frequencies.}.
The energy cut-off for the plane wave basis set was $70\,\rm Ryd$
for geometry relaxations and molecular dynamics simulations and $140\rm\,Ryd$ for computation of vibrational frequencies.
Periodic boundary conditions were applied for both crystalline solids and aqueous solutions to preserve the continuity of the media.

\begin{table*}[t]
\caption{The 
$1000(\beta-1)$, $\alpha=\beta_{\rm H_3BO_3}/\beta_{\rm H_4BO_4^-}$ and $\Delta=1000(\ln\beta_{\rm H_3BO_3}-\ln\beta_{\rm H_4BO_4^-})$
factors computed for isolated $\rm H_3BO_3$ and $\rm H_4BO_4^-$ at $T\rm=300\,K$ by \citet{Z05} using different basis sets and our result
obtained using the full normal mode spectrum and equation \ref{beta}.
The units are $\permil$.}
\label{T3}      
\centering          
\begin{tabular}{lcccc}     
\hline\hline       
& $\rm H_3BO_3$ & $\rm H_4BO_4^-$ & $\alpha=\beta_{\rm H_3BO_3}/\beta_{\rm H_4BO_4^-}$ & $\Delta=1000(\ln\beta_{\rm H_3BO_3}-\ln\beta_{\rm H_4BO_4^-})$ \\
\hline                    
 6-31+G(d)  &  216.6 &  174.4  & 1.0359  & 35.3  \\
 6-311+G(d,p) & 215.0 &  170.5  & 1.0380  & 37.3 \\
 our work & 211.3 & 167.3  & 1.0377  & 37.0 \\
\hline                  
\end{tabular}
\\
\end{table*}

The force constants and frequencies needed for the computation of the $\beta$ factors were computed using the finite displacement
scheme. Before performing the calculations of the crystal structures all atomic positions were relaxed to the equilibrium positions 
to minimize the forces acting on the atoms. We note that to compute the $\beta$ factors for crystals one formally should account for phonon dispersion.
Here we use large supercells and restrict our calculations to a single phonon wave-vector ($\Gamma$).
\citet{SCh11} has shown recently for $\rm ^{26}Mg/^{24}Mg$ fractionation in Mg-bearing minerals  
that supercells containing more than 20 atoms are sufficient to get very accurate 
$\beta$ factors even at $T=\rm 300\,K$ (error of 0.1\permil). At $T=1000\rm\,K$ the error is in the order of 0.01\permil.
The accuracy of the high temperature isotope fractionation factors computed on a single phonon wave-vector is also demonstrated
for iron-bearing minerals by \citet{BP09} and confirmed with good agreement of the predicted with the measured 
Li isotope fractionation factors between staurolite, spodumene, micas and aqueous fluid presented in our previous work \citep{KJ11}. 

Prior to the computation of the force constants and frequencies
of boron atoms in the fluids the positions of all the atoms constituting the boron-carrying molecule ($\rm H_3BO_3$ or $\rm H_4BO_4^-$) were relaxed to the equilibrium positions,
while all other atomic positions remained unchanged. The full normal mode analyzes were performed using the same method, but displacing all the atoms constituting the considered system. 
In the latter case the frequencies were obtained through the diagonalization of the full dynamical matrix \citep{S04} as implemented in CPMD code. 
The effect of the various approximations on the derived fractionation factors was studied by additional computations of 
$\rm H_3BO_3$ and $\rm H_4BO_4^-$ isolated clusters.
For that purpose we used a large, isolated simulation box with a cell length of $16\rm\,\AA$, forcing the charge density to be zero at the boundary, 
as implemented in CPMD code. In order to compute the $\beta$ factors of boron species in the aqueous fluid we apply the same method as in our recent work on Li isotopes \citep{KJ11},
with the exception that we use the pseudofrequencies, i.e. the frequencies obtained from the three force constants acting on the fractionating element, 
and formula \ref{beta} for calculation of $\beta$ factors, as discussed in section \ref{SAA}.
In order to fully account for the spatial continuity of the fluid and its dynamical motion we produced $10\rm\,ps$ 
long molecular dynamics trajectories of systems consisting of 64 H$_2$O molecules and one $\rm H_3BO_3$ or $\rm H_4BO_4^-$ molecule 
for different $T=1000\rm\,K$, $800\rm\,K$ and $600\rm\,K$ and pressure of $0.5\rm\,GPa$, which closely resembles the experimental conditions of \citet{WM05} and \citet{MW08}. 
The corresponding simulation box length is 
$13.75\rm\,\AA$ at $T=1000\rm\,K$. The $\beta$ factors were computed on the ionic configuration snapshots extracted uniformly in $0.1\rm\,ps$
intervals along the molecular dynamics trajectories. 

\subsection{Error estimation technique\label{ET}}

The errors in the computed value of the $(\beta-1)$ and $\Delta$ fractionation factors
were estimated from an average error of vibrational frequencies computed using the chosen
DFT method. \citet{FS95}, \citet{MT02} and \citet{AZZ10} estimated the errors made in calculations of vibrational frequencies
of small molecules using different DFT functionals. 
According to these works the BLYP functional systematically overestimates the harmonic frequencies by $\sim3.5\,\%$,
with a deviation from the mean offset of $\sim1\,\%$. 
Therefore, we expect that using BLYP functional 
the $(\beta-1)$ and $\Delta$ values are systematically overestimated by $\rm 7\,\%$
and that in addition there is a $\rm 2\,\%$ error in derived $(\beta-1)$ factors.
Similar errors result from using other functionals or even more sophisticated and time consuming post-Hartree-Fock
methods such as MP2 \citep{FS95,AZZ10}.

\section{Results and discussion}

\subsection{Test of the computational method \label{TEST}}

First, we illustrate the performance of the approximation proposed in section \ref{SAA} by computing the $\beta$ factors
for the isolated $\rm H_3BO_3$ and $\rm H_4BO_4^-$ molecules and selected crystalline solids. In Figure \ref{F1} we present three sets of calculations of $\beta$ factors: 
(1) the ``exact" result obtained from a full normal mode analysis and formula \ref{beta}, (2) the results obtained 
applying \citet{KJ11} method based on Eq. \ref{BAPX}, (3) the results obtained using pseudofrequencies computed for the fractionating element and 
Eq. \ref{beta} for the estimation of the $\beta$ factor. The numerical values for selected temperatures are reported in Table \ref{T1}.
Approach (3) results in much better agreement with the ``exact" result. For $\rm H_3BO_3$, the $\beta$ factor 
is overestimated by only $0.5\permil$ and $1.5\permil$ for temperatures of $\rm 800\,K$ and $\rm 600\,K$ respectively. Applying method (2),
the error is more pronounced, $2.2\permil$ and $6.7\permil$ respectively.
In the case of molecular $\rm H_4BO_4^-$, the errors using method (3) for the same temperatures are only $0.1\permil$ and $0.5\permil$ respectively.
The same behavior is shown for dravite and boromuscovite crystalline solids that contain boron in the coordinative arrangement that resemble
the configurations of aforementioned B-bearing molecules.
For $T>600\rm \,K$ the proposed method represents only a few percent correction to the
approximation given by equation (\ref{BAPX}), so the relation (\ref{TESTP}) is satisfied.
It is evident that for B-bearing materials
considered here the improvement made by using the pseudofrequencies based approach is substantial. It corrects for about 75$\%$ of error of the \citet{BM47} 
approximation (Eq. \ref{BAPX}). However, the question of general applicability of the proposed method to other isotopic systems would require careful testing 
on a large set of materials, which is well beyond the scope of the current paper.

\subsection{B isotope fractionation in gas and fluid phases}

\subsubsection{B isotope fractionation between $\rm H_3BO_3$ and $\rm H_4BO_4^-$ in the gas phase}

In a first step of our investigation of boron-rich aqueous fluids we derived the full frequency spectra of molecules in the
gas phase (isolated molecules). The relevant $\beta$ factors were computed using Equation \ref{beta}. These studies were performed
in order to compare our results with the published values of \citet{Z05}, both computed using the same DFT BLYP functional.
In Table \ref{T1a} we report the computed frequencies that are affected by the different B isotope substitutions along with other 
theoretical estimations and experimental measurements. The computed frequencies are in good agreement with earlier theoretical predictions and
show similar agreement with the experimental measurements. 
The results in terms of computed $\beta$ factors for the two considered species are reported in Figure \ref{F6}, where we compare our results with 
the values computed using frequencies of \citet{Z05}.
The comparison of the two sets of calculations reveals that our $\beta$ factors for both species are smaller by $\sim1\,\permil$ at $\rm 600\, K-1000\,K$
than the values of \citet{Z05}.
However, the difference between $\beta$ factors of $\rm H_3BO_3$ and $\rm H_4BO_4^-$ remains nearly identical in both sets of calculations 
and the agreement is nearly perfect for higher temperatures.
We note, that for the comparison we used the frequencies of \citet{Z05} computed using 6-31+G(d) 
basis set, as only these are provided by the authors. $\beta$ and $\alpha$ factors obtained at $T=300\rm\,K$ using a more extended
6-311+G(d,p) basis set indicate that the $\beta$ factors using 6-31+G(d) basis set are not fully converged. In particular, the $(\beta-1)$ factor of
$\rm H_4BO_4^-$ computed with 6-311+G(d,p) basis set is $\rm 3.9\,\permil$ smaller than the one derived using 6-31+G(d).
In Table \ref{T3}, we compare these results with the results of our calculation. It is clearly seen that for lower temperatures such as $T=300\rm\,K$ 
the values computed with 6-311+G(d,p) basis set are in better agreement with our results indicating that plane-wave based DFT approach 
we use provides adequate vibrational frequencies and resulting isotope fractionation factors.

\begin{figure*}[t]
\includegraphics[angle=270,width=3.5in]{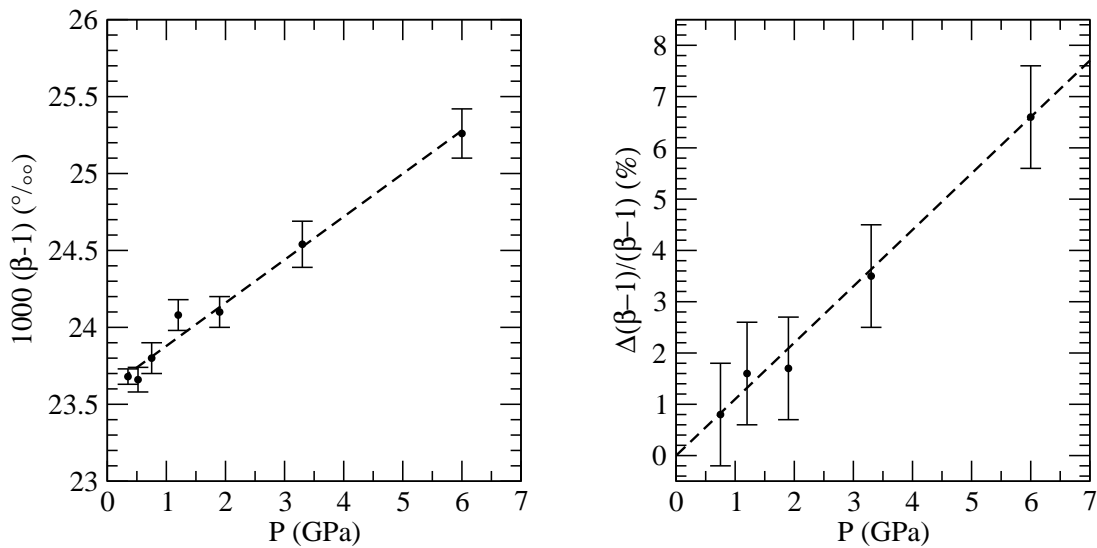}
\caption{Left panel: pressure dependence of the $\beta$ factor of neutral fluid ($\rm H_3BO_3$) at $T\rm=1000\,K$. 
Dashed line is the linear regression fit to the calculated values (points): $1000(\beta-1)=23.6+0.28\,P{\rm (GPa)}$; Right panel: The computed change of the $\beta$ factor with pressure (symbols) 
in comparison to the increase in the $\beta$ factor derived from the frequency shifts of the $666\,\rm cm^{-1}$ and $1454\,\rm cm^{-1}$ lines measured by \citet{SV05} (dashed line).
The comparison is made assuming that $(\beta-1)\sim\nu^2$.
\label{F7a}}
\end{figure*}

\begin{figure*}[t]
\includegraphics[angle=270,width=3.5in]{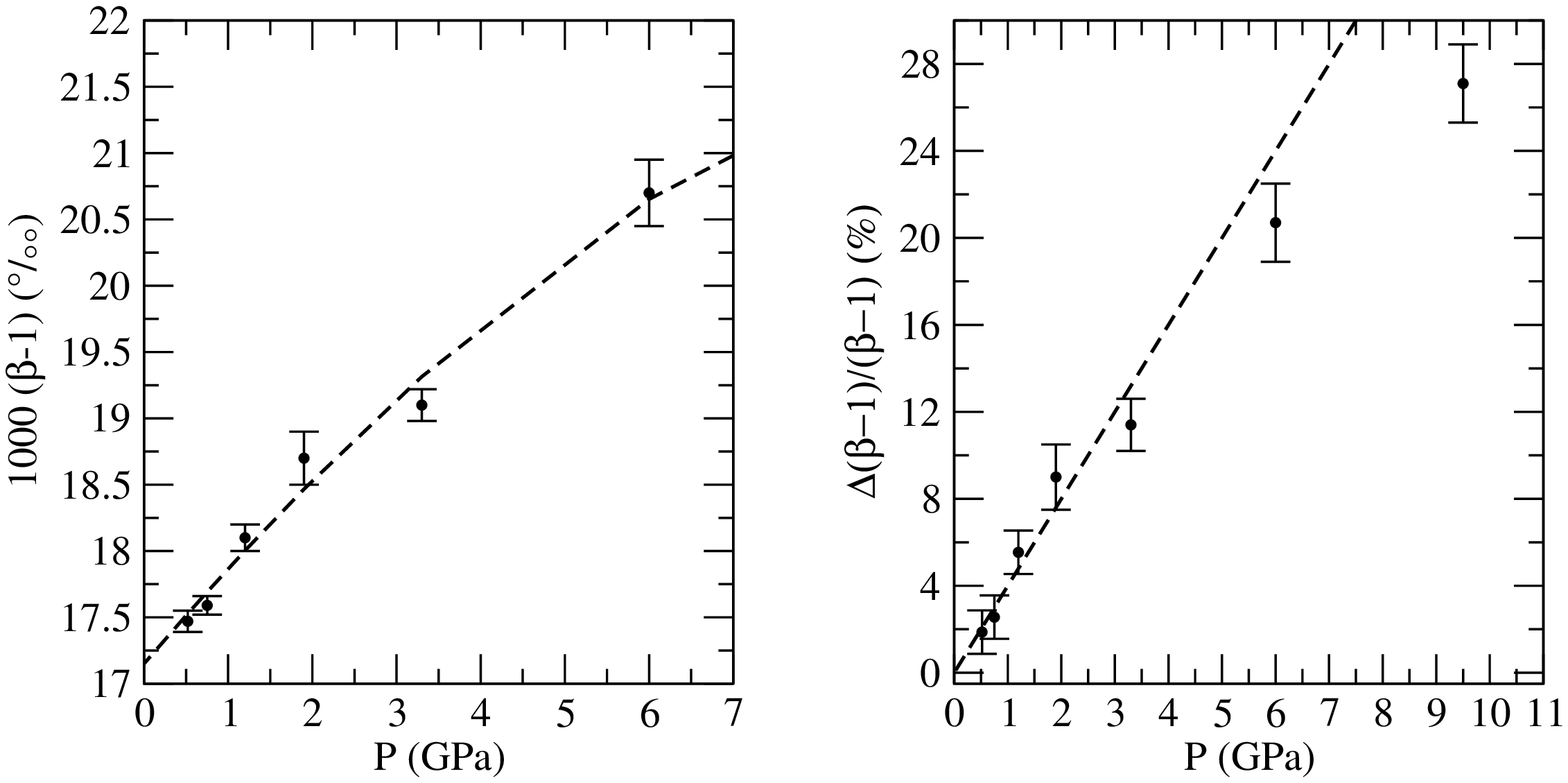}
\caption{Left panel: pressure dependence of the $\beta$ factor of strongly basic fluid ($\rm H_4BO_4^-$) at $T\rm=1000\,K$. 
Dashed line is the regression fit to the calculated values (points): $1000(\beta-1)=17.15+0.754\,P-0.027\,P^2$, where pressure is given in $\rm GPa$; Right panel: 
The computed change of the $\beta$ factor with pressure in comparison to the increase in the $\beta$ factor 
derived from the frequency shift of the $975\,\rm cm^{-1}$ line measured by \citet{SV05} (dashed line). The comparison is made assuming that $(\beta-1)\sim\nu^2$.
\label{F7b}}
\end{figure*}

\subsubsection{B isotope fractionation between $\rm H_3BO_3$ and $\rm H_4BO_4^-$ in aqueous fluid}

In order to obtain the temperature dependent $\beta$ factor for aqueous fluids
we fitted the function $1+A/T^2+B/T^4$ to the computed values using the least squares minimization procedure. 
The computed $\beta$ values for $\rm H_3BO_3$ in fluid are: $1.02366\pm0.00012$, $1.03624\pm0.00018$ and $1.06262\pm0.00010$ and for $\rm H_4BO_4^-$ 
in fluid are: $1.01745\pm0.00005$, $1.02650\pm0.00010$ and $1.04597\pm0.00015$,
for the temperatures of $1000\rm \,K$, $800\rm \,K$ and $600\rm \,K$ respectively. The resulting temperature dependent $\beta$ factor for $\rm H_3BO_3$ is $\beta=1+2.416\cdot 10^4/T^2-5.823\cdot 10^{8}/T^4$ and for $\rm H_4BO_4^-$ is 
$\beta=1+1.772\cdot 10^4/T^2-4.234\cdot 10^{8}/T^4$.

The results for $\rm H_3BO_3$ and $\rm H_4BO_4^-$ in aqueous solutions are shown in Figure \ref{F5}. As was observed for the isolated molecules, 
the $\beta$ factor of $\rm H_3BO_3$-bearing fluid is substantially larger than the one for the $\rm H_4BO_4^-$.
This can be understood in terms of the substantial difference in the B-O bond lengths exhibited by the two considered species. In case of isolated molecules our calculations
indicate a B-O bond length of $1.40\rm\,\AA$ for $\rm H_3BO_3$ and $1.51\rm\,\AA$ for $\rm H_4BO_4^-$.
We compared our $\beta$ factors with the values computed by \citet{SV05}, which were derived by the combination of force field methods and experimental data to derive accurate vibrational frequencies. 
For $\rm H_3BO_3$ we got a nearly identical result. In case of $\rm H_4BO_4^-$
our calculation predicts a value which is lower by $\rm 2-4\,\permil$. However, \citet{RB10} and \citet{RB07} revealed the improper assignment of a major fractionating vibrational mode
of $\rm H_4BO_4^-$ in the force field by \citet{SV05}. This leads to the underestimation of the fractionation factor between aqueous $\rm H_3BO_3$ and $\rm H_4BO_4^-$
by \citet{SV05}.
Assuming that $\alpha\propto T^{-2}$ and having the difference between BLYP calculations of \citet{RB10} and \citet{SV05} of $\Delta\alpha=16.4\,\permil$ at $T\rm=300\,K$,
the value reported by \citet{SV05} should be underestimated by $\Delta\alpha=16.4\cdot 2(300/T)^2\,\permil$, which results
in $\Delta\alpha\sim1.5\,\permil$ at $T\rm=1000\,K$. Corrected in such as way result of \citet{SV05} is also plotted in Figure \ref{F5}. 
It is now very consistent with our prediction.

\subsubsection{Discussion of computational errors}

Most previous computational studies of boron isotope fractionation in aqueous solutions concentrate on the computation of the isotope fractionation 
at ambient conditions \citep{RB10,RB07,LT05,Z05}. \citet{RB10} performed detailed analysis of impact of the chosen computational method (HF, MP2, different DFT functionals) 
and size of the basis set on the calculated fractionation factors between $\rm H_3BO_3$ and $\rm H_4BO_4^-$. 
They found that DFT methods are not performing well for the borate system and concluded that DFT {\it ''is of limited usefulness in chemically 
accurate predictions of isotope fractionation in aqueous systems"} \citep{RB10}.
The empirically derived error of the derived fractionation factor is of the order of $5-10\,\permil$ for a total fractionation of $\sim30\rm \,\permil$. 
We note that this is expected and clearly visible if we apply the error estimation procedure outlined in section \ref{ET}. For instance, at room temperature 
the derived beta factors using the BLYP functional are $213.6\,\permil$ and $173.3\,\permil$ respectively (\citet{RB10}, Table 2). This gives a fractionation 
factor of 1.0343. Following our error estimation scheme, the absolute error of the fractionation factor is $10.5\,\permil$, and the properly reported computed value 
is $\alpha=1.034\pm 0.011$. When one corrects for
the systematic error of $7\%$ and assumes $2\%$ of statistical error on $\beta$ factors, then the value of $\alpha$ decreases and the 
error is slightly smaller, i.e. $\alpha=1.032\pm 0.008$.
This is in good agreement with the experimental data reported in \citet{RB10} and explains the spread of the values computed using different methods
and reported in that paper. It is very difficult to get the fractionation factors for ambient conditions, as the fractionation factor is often just a small fraction of
the relevant $(\beta-1)$ factors, $(\alpha-1)\sim0.15(\beta-1)$ in the considered case. Assuming that $(\alpha-1)=0.15(\beta-1)$, a $2\%$ error in the $(\beta-1)$ factors 
leads to an absolute error in $(\alpha-1)$ of $0.04(\beta-1)=0.04(\alpha-1)/0.15\sim0.27(\alpha-1)$, i.e. $\sim 27\%$ of relative error 
in the derived fractionation factor $(\alpha-1)$. On the other hand, 
we note that such a big error is not substantially larger than the uncertainties in the experimental data reported by \citet{RB10} in their Figure 2.
Thus, the case of boron fractionation in aqueous fluid at ambient conditions does not necessarily show the limited usefulness of DFT
in the prediction of isotope fractionation factors, but only reflects the fact that precise estimation or measurement of the B isotopes fractionation factors
at ambient condition requires unprecedented accuracy of both experimental or computational techniques. For instance, in order to get the value of $(\alpha-1)$ 
with a relative error of $5\%$ (at ambient conditions) one needs to estimate the $(\beta-1)$ factors or measure relevant quantities with precision of less than $1\%$.
At higher temperatures the situation is different.
Looking just at the fractionation between $\rm H_3BO_3$ and $\rm H_4BO_4^-$ in the gas phase or the aqueous solution one can see that for $T>600\rm\,K$ the
fractionation factor between the two substances, $(\alpha-1)$, is at least $25\%$ of the $(\beta-1)$ factor. This results in smaller $0.04/0.25\sim 16\%$ for $T=600\rm\,K$
and $0.04/0.36\sim 11\%$ for $T=1000\rm\,K$
relative error, which is acceptable in our calculations. Nevertheless, this case shows the importance of proper error estimation on the computed
fractionation factors. Such an estimation is usually omitted or not provided explicitly, which can lead to wrong conclusions when the theoretical prediction
is confronted with the measured data.

\subsubsection{Pressure dependence of the fluid fractionation factor}

In our recent paper \citep{KJ11} we have shown that due to compression  the $\beta$ factor of Li in aqueous fluid increases with increase
in pressure (for $P\rm>2\rm \,GPa$). The same should happen for $\rm H_3BO_3$ and $\rm H_4BO_4^-$ aqueous fluid as the vibrational frequencies 
of boron species in aqueous fluid increase with increase in pressure \citep{SV05,STH05}.
Having the experimental data we checked whether the derived pressure-dependent $\beta$ factors are consistent with the 
pressure shifts of vibrational frequencies of considered boron species measured by \citet{SV05}. 
For that purpose we performed a set of calculations using supercells containing 8 water molecules and the relevant boron species.
We note that in line with our previous results for Li \citep{KJ11}, the obtained values of ($\beta-1$) at $P\rm=0.5\,GPa$ are within $\rm 0.1\,\permil$ 
in agreement with the values obtained for supercells containing 64 water molecules.
The results are given in Figures \ref{F7a} and \ref{F7b}. The computed $(\beta-1)$
values for $\rm H_3BO_3$ fluid show a linear dependence in pressure, $(\beta-1)=23.60+0.28\,P\rm(GPa)\, \permil$. This is expected as $\rm (\beta-1)\propto\nu^2\sim\nu_0^2+2\nu_0\Delta\nu$ \citep{S04}
and $\Delta\nu$ is a linear function of pressure \citep{SV05,STH05}. In case of $\rm H_4BO_4^-$ the pressure-dependence is linear up to $P\rm \sim2-3\,GPa$ and it becomes
less steep at higher pressures. In order to quantitatively check the consistency of our prediction with the measured vibrational frequency shifts
of \citet{SV05} we derived the relative shifts in the $(\beta-1)$ factor assuming that $\rm (\beta-1)\propto\nu^2$
and the measured pressure dependence of the frequency shifts: $\Delta \nu=2.15{\rm\,cm^{-1}}\cdot P(\rm GPa)$ and $\Delta \nu=3.50{\rm\,cm^{-1}}\cdot P(\rm GPa)$ for $\rm 1454\,cm^{-1}$ 
and $\rm 666\,cm^{-1}$ vibrational frequencies of $\rm H_3BO_3$ and $\Delta \nu=6.47{\rm\,cm^{-1}}\cdot P(\rm GPa)$ for the $\rm 975\,cm^{-1}$ vibrational frequency of $\rm H_4BO_4^-$. 
The chosen vibrational frequencies are the ones affected by the different B isotope substitution.
Our predicted shift of $(\beta-1)$ matches well the shifts derived from the measured frequency shifts.
Such a good agreement with the experimental data validates further our computational approach and shows that {\it ab initio} calculations can be successfully used 
in derivation of the pressure dependence of the fractionation factors and pressure-induced vibrational frequencies shifts. Moreover, first principles calculations 
can be useful in extrapolation of the experimental values for $\beta$ and $\Delta\nu$ to more extreme conditions, which otherwise are extremely difficult to reach by experimental 
techniques.

\subsection{Fluid-mineral fractionation}

Next we present the results of the fractionation between boron bearing fluids and minerals such as dravite, olenite and boromuscovite.
The aim of these studies is to investigate the mechanisms driving the fractionation process, the role of coordination and the B-O
bond length. Below we discuss each case separately.

\begin{figure}[t]
\includegraphics[angle=270,width=2.0in]{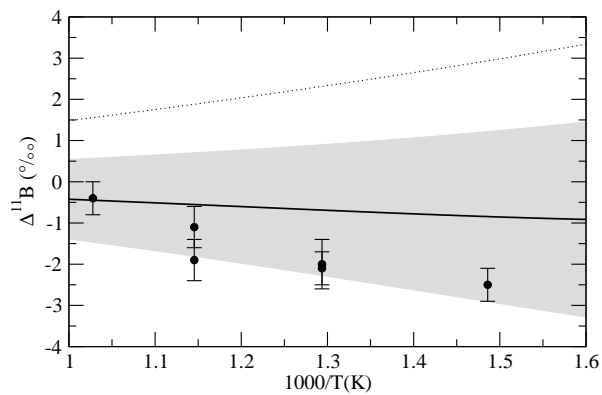}
\caption{The fractionation factors between tourmaline and aqueous fluid. The solid line represents our prediction for the fractionation between
dravite and $\rm H_3BO_3$ neutral fluid and the shadowed region represent the computational uncertainty. The dotted line represents our prediction for the fractionation between olenite 
containing $\rm B^{[3]}$ species only and neutral fluid.
The computational error is similar in both cases. The data points are the values measured for dravite-fluid system by \citet{MW08}.
\label{F3}}
\end{figure}

\subsubsection{Tourmaline-neutral fluid}

\citet{MW08} measured the boron isotope fractionation between tourmaline and neutral fluid at $T=400-700\,\rm ^{o}C$
and $P\rm=0.2\,GPa$. In the experiment the tourmaline was represented by dravite.
In contrast to the former measurements of \citet{P92} the measured fractionation is very small
and does not exceed $2.5\,\permil$ at $400\rm\,^{o}C$. Our calculated fractionation curve together with the experimental 
data are given in Figure \ref{F3}. Our result correctly reproduces the experimental measurements within the computational accuracy.
The dravite-fluid fractionation is small as the two materials contain boron in $\rm BO_3$ units.
We also predict a small fractionation between olenite carrying 3-fold coordinated boron only and aqueous fluid, although the olenite-fluid fractionation is positive
because of the shorter B-O bond lengths for olenite ($\rm 1.378\,\AA$ vs. $\rm 1.397\,\AA$).

\begin{figure}[t]
\includegraphics[angle=270,width=2.0in]{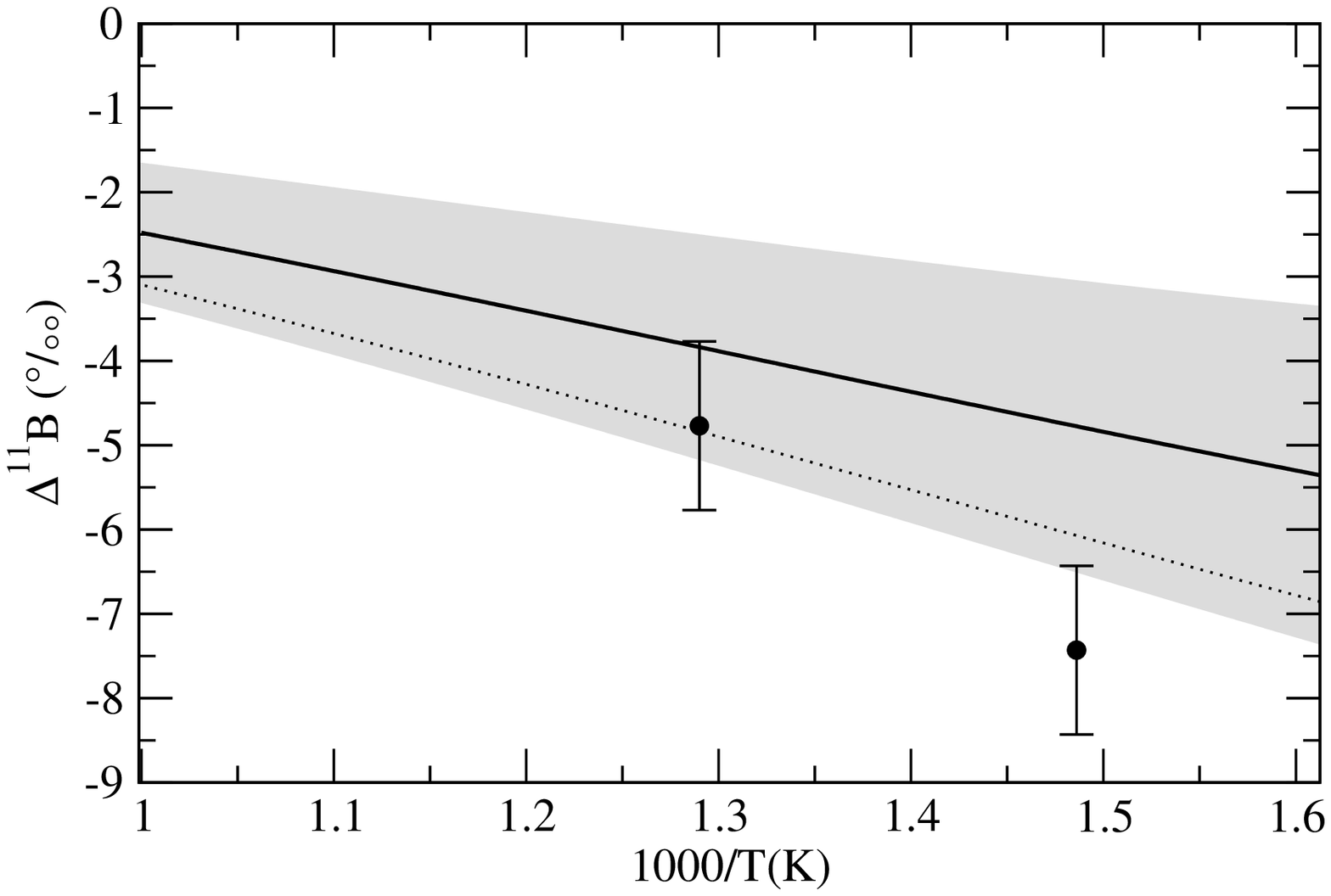}
\caption{The fractionation factors between boromuscovite and basic aqueous fluid (fluid containing $\rm H_4BO_4^-$). The lines represent our results
assuming the presence of boron species in form of $\rm H_4BO_4^-$ only (solid line) and mixture of $90\%$ of $\rm H_4BO_4^-$ and $10\%$ of $\rm H_3BO_3$ 
(dotted line) in the fluid.
The data points are the values measured by \citet{WM05}. 
The uncertainty of calculated values is indicated by shadowed area and is similar for both curves.
\label{F4}}
\end{figure}

\subsubsection{Boromuscovite-strongly basic fluid \label{SB}}

Boromuscovite synthesized in the experiments of \citet{WM05} consisted of two type of polytypes, 1M ($\sim10\,\%$) and 2M1 ($\sim90\,\%$).
In order to be consistent with the experimental conditions, we derived the $\beta$ factors for both polytypes and computed their weighted 
average. We note that the $\beta$ factors for both polytypes of mica are similar with the difference in $(\beta-1)$ not larger than $3\rm\,\%$. 
This is consistent with similar B-O bond lengths ($1.532\rm\,\AA$) found in both polytypes.
Boromuscovite contains boron in tetrahedral sites.
Therefore, in order to investigate the impact of the B-O bond length on the fractionation 
we first compare the fractionation between the mineral and a strongly basic fluid containing boron in $\rm H_4BO_4^-$.
The result, together with the measurements of \citet{WM05} of the fractionation between boromuscovite and strongly basic fluid,
is summarized in Figure \ref{F4}. Our calculations predict a negative fractionation between mica and the $\rm H_4BO_4^-$ fluid.
The agreement of our prediction with the experimental measurements is relatively good; however, the experimental data
indicate slightly stronger fractionation. We note that the experimental conditions of \citet{WM05}
do not assure that the measured basic fluid contained four-fold coordinated boron species, i.e. $\rm H_4BO_4^-$, only.
As we indicate in the Figure \ref{F4}, the presence of as little as $10\%$ of $\rm H_3BO_3$ in the measured basic fluid 
brings the prediction and measurements into much better agreement. It makes sense that the $\beta$ factor of boromuscovite is smaller than for aqueous $\rm H_4BO_4^-$
as the average B-O bond length in mica is $\rm 1.532\,\AA$, while it is $\rm 1.513\,\AA$ and therefore shorter in case of aqueous $\rm H_4BO_4^-$.

We note that in our derivation we assumed that the fluid consists mostly of $^{[4]}\rm B$ species. As we already mentioned, 
although \citet{WM05} call the fluid ``strongly basic" its exact composition, especially the amount of $^{[3]}\rm B$ species is unknown.
However, if for instance the $^{[4]}\rm B$ to $^{[3]}\rm B$ ratio was $1$, then the predicted mica-basic fluid fractionation at $\rm 800\,K$ would be $5\,\permil$
larger than measured. This would result in a large inconsistency between the computed values and the experimental data.
On the other hand, good agreement between the prediction and the measurements
indicates that the strongly basic fluid was dominated by $\rm H_4BO_4^-$ species, which is in line with previous studies \citep{Z05,SV05}.

\begin{figure}[t]
\includegraphics[angle=270,width=2.0in]{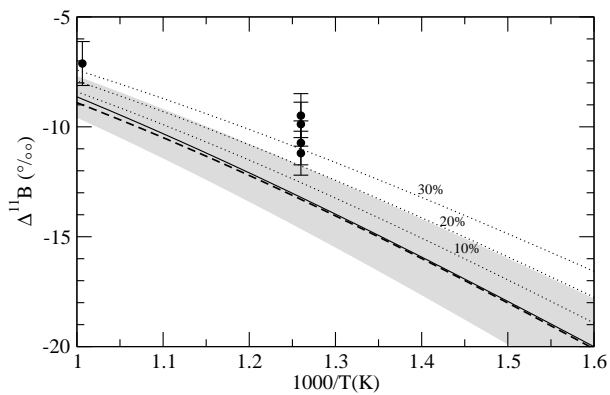}
\caption{The fractionation factors between boromuscovite and aqueous fluid.
The data points are the values measured by \citet{WM05}. 
The solid line represents the result obtained for ambient pressure.
The dashed line represents the result for fluid containing $\rm H_3BO_3$ only obtained for $P=3\,\rm GPa$ accounting for compression and thermal expansion, 
with the uncertainties in calculated values indicated by shadowed area.
The dotted lines represent the results assuming different admixture of four-fold coordinated boron species (represented by $\rm H_4BO_4^-$ with abundance indicated in the figure) to the fluid.
The computational error is comparable for all the results.
\label{F8}}
\end{figure}

\subsubsection{Boromuscovite-neutral fluid \label{BNF}}

The fractionation between boromuscovite and neutral fluid involves a change in coordination from $\rm B^{[4]}$ in boromuscovite 
to $\rm B^{[3]}$ in neutral fluid. \citet{WM05} measured the fractionation between the two materials at $\rm 3\,GPa$, shown
in Figure \ref{F8}. The predicted fractionation is about $3\pm2.5\,\permil$ larger than
the measured value. Looking for potential sources of this discrepancy, we have checked for the effect of the change in lattice parameters
due to combined thermal expansion and compression. For that purpose we applied the EOS of \citet{HP11} for muscovite, which 
gives $4.4\%$, $3.8\%$ and $3.1\%$ decrease in volume for $T=600\,\rm K$, $800\,\rm K$ and $1000\,\rm K$ respectively and $P=3\rm\,GPa$.
As muscovites show highly anisotropic compressibility patterns, in line with \citet{CZ95} we applied the $T$ and $P$ driven change in volume 
assuming the $16\%$, $19\%$ and $65\%$ contribution to compression along the $\vec{a}$, $\vec{b}$, $\vec{c}$ lattice vectors.
On the other hand $(\beta-1)$ factors of $\rm H_3BO_3$ and $\rm H_4BO_4^-$ in aqueous solutions at $P\rm=3\,GPa$ increase 
by $3.5\%$ and $11.8\%$ respectively (Figures \ref{F7a} and \ref{F7b}), leading to pressure-induced increase in the boromuscovite-aqueous fluid fractionation
at the experimental pressure. The $\Delta$ factor, corrected for effects of thermal expansion and compression of boromuscovite and compression of fluid,
is also given in Figure \ref{F8}. Because the high $T$ and $P$ effects result in similar increases in the $\beta$ factors for both solid and aqueous fluid,
the resulting fractionation factor between these two phases is close to the one derived at ambient conditions.
Therefore, thermal expansion and compression effects cannot explain the observed discrepancy between prediction and the measurements of \citet{WM05}.

On the other hand, the comparison of our results with the experimental data suggests that the fractionation between boromuscovite and fluid is the same as between
$\rm H_3BO_3$ and $\rm H_4BO_4^-$ fluids (see Figure \ref{F5}), which is at odds with the non-negligible and negative fractionation between boromuscovite and
a strongly basic fluid. In section \ref{SB} we have shown that we are able to correctly reproduce the
fractionation between boromuscovite and strongly basic fluid, which indicates that our result for boromuscovite is reliable.
This suggests that another, unaccounted effect leads to the decrease of the boron isotope fractionation between mica and neutral fluid 
in the experiments of \citet{WM05}.

One possible solution for the discrepancy 
is a non-negligible amount of boron residing in four-fold coordinated configurations in neutral solution. This is in line with the Raman spectroscopy measurements
of \citet{STH05}, who detected a broad peak in the Raman spectra of neutral $\rm H_3BO_3$-dominated fluid and attributed it to $\rm B^{[4]}$ species. 
The integrated area of this peak, compared to the peak of the Raman $877\,\rm cm^{-1}$ line
of $\rm B^{[3]}$ species, indicates the presence of at least $15-30\,\%$ of $\rm B^{[4]}$ species by mole fraction. Assuming that there is $\rm 15-30\,\%$
of $\rm B^{[4]}$ species present in the fluid and that the $\beta$ factor of these species is similar to that of $\rm H_4BO_4^-$, the fractionation factor
between boromuscovite and $\rm H_3BO_3$ aqueous fluid decreases bringing the theory and the experiment to better agreement, which is illustrated in Figure \ref{F8}. 
If this interpretation is true, it suggests that boron isotope fractionation could be used to gather information on the speciation of B in aqueous fluids.

\begin{figure}[t]
\includegraphics[angle=270,width=2.0in]{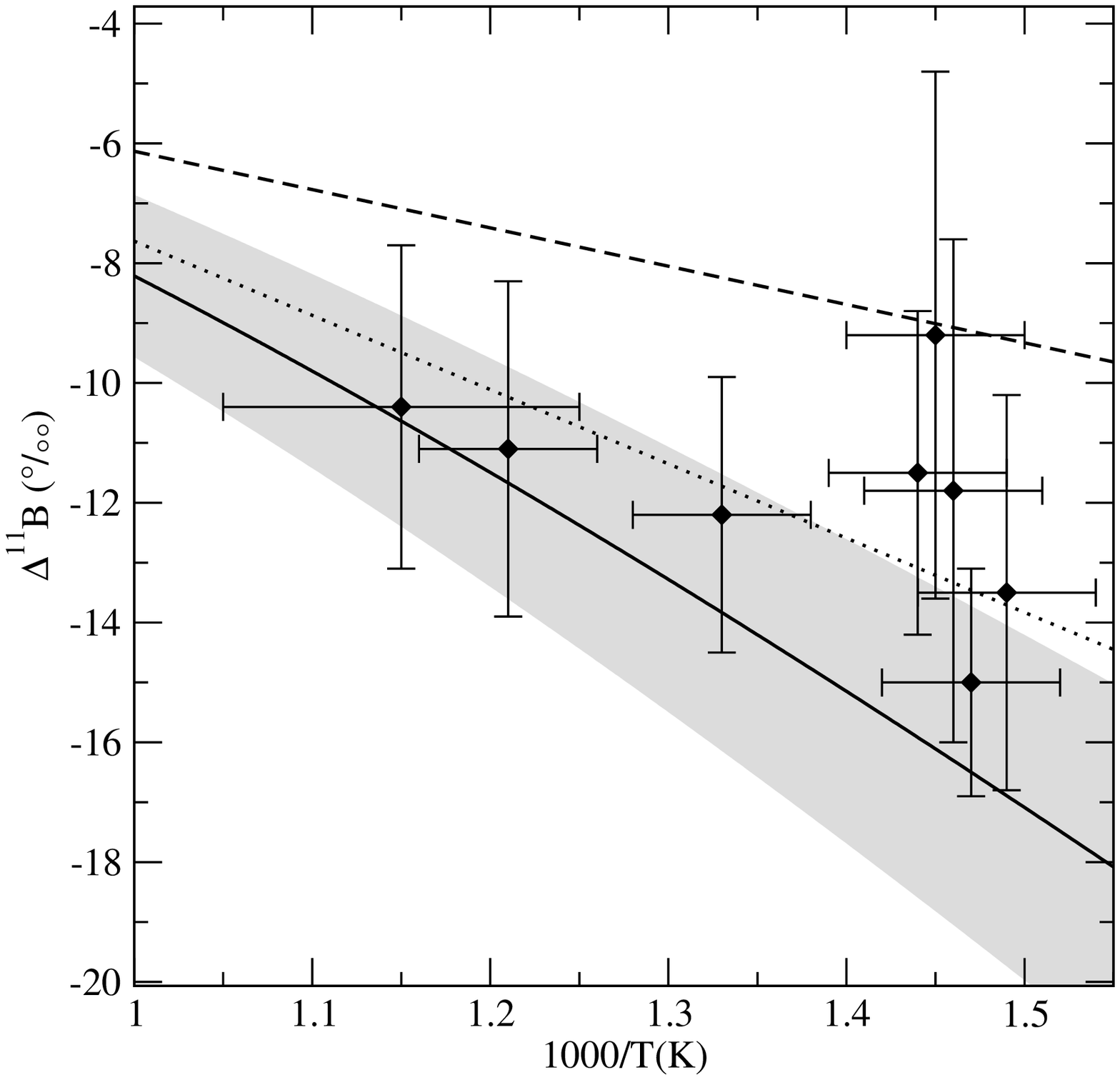}
\caption{ The fractionation factors between mica and tourmaline. The solid line represents our value for fractionation between B-muscovite
and dravite. The dashed line is the experimental fractionation factor between tourmaline and mica determined by \citet{WM05} and \citet{MW08}. 
The experimental error
is $\rm 2\,\permil$.
The dotted line is the experimental fractionation factor of \citet{WM05} and \citet{MW08} but corrected for the presence of $\rm B^{[4]}$ species
in the neutral fluid in the high $P$ experiments of \citet{WM05}, as is discussed in the text.
The diamonds are the data from natural samples taken from \citet{KM11} and references herein.
The uncertainties in calculated values are indicated by shadowed area.
\label{F2}}
\end{figure}

\subsection{B isotope fractionation between minerals}

The boron isotope fractionation between B-bearing crystalline solids has received considerable attention recently \citep{WM05,MW08,KM11,M05,H02}.
We focus here on the investigation of boron isotope fractionation between mica and tourmaline as boron atoms in these minerals 
occupy sites of different coordination, which should result in a large B isotope fractionation between
these two minerals. In micas boron substitutes for silicon in the four-fold coordinated site \citep{WM05}, 
while in tourmaline (dravite) it is incorporated in the three fold coordinated site \citep{MW08}. Comparing the B isotope fractionation 
between different minerals, melts and fluids \citet{WM05} have shown that the fractionation between two materials of different 
B coordination is large, reaching $5\,\permil$ at $\rm 1000\,K$ and much higher values at lower temperatures.
\citet{KM11},\citet{M05} and \citet{H02} measured the fractionation between coexisting phases of the two minerals in natural samples.
The fractionation between these two minerals is also derived from experimental isotopic fractionation data of B-muscovite-fluid \citep{WM05}
and tourmaline-fluid \citep{MW08} systems. The results of these measurements and our computed $T$-dependent fractionation curve are given in Figure \ref{F2}.
The first striking observation is that our predicted fractionation factors are much larger (taking the absolute value) than 
the experimental values \citep{WM05,MW08}. The latter are also inconsistent with the natural samples data of \citet{KM11} and
previous studies discussed in that paper \citep{M05,H02}. On the other hand, the measurements on natural samples are consistent with our calculated values,
which tends to validate our predictions.
We notice that the most recent measurements of boron isotope signatures of tourmaline and white mica from the Broken Hill area in Australia by \citet{KM11} 
indicate for the assumed temperature of 600 $^{\rm o}$C that the fractionation factor between the two phases is 10.4$\pm2.7\,\permil$, which is in good agreement with
our computed value of $10.7\pm1.8 \, \permil$. 
The experimental mica--tourmaline B isotope fractionation factors of \citet{WM05} and \citet{MW08} are $2\,\rm\permil$ and $6\,\rm\permil$ smaller at temperatures of $1000\,\rm K$ and $800\,\rm K$
respectively, with an experimental uncertainty of $2\rm\,\permil$. However, this discrepancy can be resolved by assuming that in the experiments of \citet{WM05} the fluid 
contained a significant admixture of $\rm B^{\rm [4]}$ species, which leads to the underestimation of the experimental boromuscovite-fluid fractionation factor
by $\sim2\,\permil$ at $1000\,\rm K$ and $\sim3.5\,\permil$ at $1000\,\rm K$, as is seen in Figure \ref{F8}. The experimental mica--tourmaline fractionation factor corrected 
for the presence of $\rm B^{\rm [4]}$ species is also plotted in Figure \ref{F2}. It is now more consistent with the natural data. 
We note that this result independently supports the conclusion underlined in section \ref{BNF} and result of \citet{STH05}
that high-$P$, B-bearing neutral fluids contain significant admixtures of $\rm B^{\rm [4]}$ species.

Olenite is a mineral which can incorporate boron in both trigonal and tetrahedral sites as it substitutes for both Al and Si atoms.
It is therefore interesting to check the fractionation of boron isotopes between the two differently coordinated sites in one mineral
and compare it with the above result for mica and tourmaline.
The computed fractionation between the trigonal and tetrahedral sites at 600 $^{\rm o}$C is $10.6\pm1.9\,\permil$, which is consistent with 
the fractionation between tourmaline and mica,
indicating that the coordination of the B atom is the driving factor for the fractionation of the B isotopes.
Similarly, we computed the boron isotopes fractionation between trigonal and tetragonal boron sites in dravite.
In order to create the tetragonal B site we replaced one Si atom with B and we added one H atom forming an additional OH group to compensate the charge.
The computed fractionation between the sites at 600 $^{\rm o}$C is $8.9\pm1.7\,\permil$, which is also in agreement with the aforementioned results.
Next, we will show that the value of the $\beta$ factor depends not only on coordination but is also strongly correlated with B-O bond length.

\begin{figure}[t]
\includegraphics[angle=270,width=2.0in]{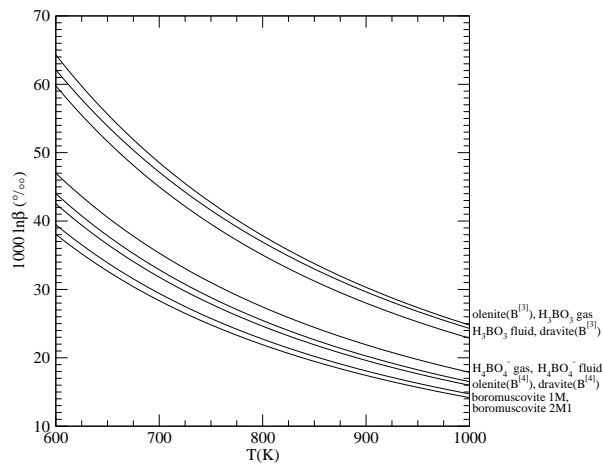}
\caption{$\beta$ factors for various considered materials as a function of temperature. The materials are indicated on the right side. Their
order reflects the value of the $\beta$ factor at $1000\,\rm K$ from the largest (top) to the smallest (bottom). \label{F10}}
\end{figure}

\subsection{
Fractionation between $\rm B^{[3]}$ and $\rm B^{[4]}$ materials} 

The $\beta$ factors computed for all the considered materials are grouped together in Figure \ref{F10}.
$\beta$ factors can be grouped into two sets, one that includes materials with boron
in three-fold coordination and another one that includes materials having boron
in four-fold coordination. For olenite and dravite we also computed the $\beta$ factors with boron sitting on four-fold coordinated site.
The $\beta$ factor for these crystalline solids with given $\rm B^{[3]}/B^{[4]}$ ratio can be derived as a weighted average of the 
$\beta$ factors obtained for boron sitting on the two differently coordinated sites. The fractionation factor
between materials of different boron coordination is $\sim8\,\permil$ on average at $T=1000\,\rm K$. We note that it is $\sim3\,\permil$ larger than the one deduced by \citet{WM05} 
from measurements performed on solids, silicate melts and fluids, but this can be attributed to the underestimation of the fractionation factors
for boromuscovite-fluid system by \citet{WM05} due to potential admixture of $\rm B^{[4]}$ species in the investigated fluid.

\begin{figure}[t]
\includegraphics[angle=270,width=2.0in]{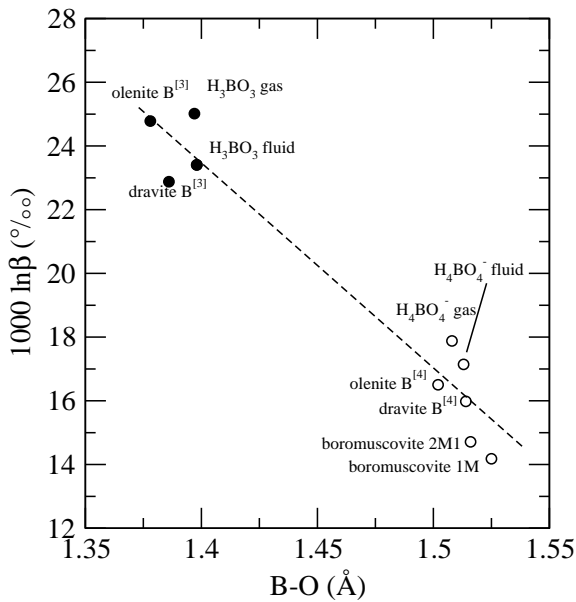}
\caption{The $\beta$ factor at $T\rm=1000\,K$ for various considered materials as a function of B-O bond length. Filled circles represent the values obtained for boron in trigonal sites and 
open circles represent the values obtained for tetragonal sites. \label{F7}}
\end{figure}

Our results show also a substantial spread of $\beta$ factors of substances containing boron of a given coordination.
The spread is at least $4\,\permil$ and results from different B-O bond lengths.
We illustrate this in Figure \ref{F7} by plotting together the $\beta$ factors derived for all considered materials 
at $T=1000\rm\,K$ as a function of B-O bond length. It is clearly seen that there is a roughly linear correlation between the $\beta$ factor and B-O bond length,
which is especially evident comparing the results for crystalline solids.
For instance, out of the considered $\rm B^{[4]}$-bearing minerals boromuscovite has the longest B-O bond length of $\rm 1.525\,\AA$ (1M) and $\rm 1.516\,\AA$ (2M1), 
followed by dravite $\rm 1.514\,\AA$ and olenite with B-O bond length of $\rm 1.502\,\AA$.
This tracks the differences in the $\beta$ factors derived for these materials.
In addition the materials having B$^{[3]}$ species only exhibit shorter bond lengths of $\rm \sim1.37\,\AA$ and higher $\beta$ factors, while the materials 
containing B$^{[4]}$ species having bond lengths of about $\rm \sim1.52\,\AA$ show much smaller $\beta$ factors.
Therefore, the tighter bonding of B$^{[3]}$ species likely explains why the heavy B isotope prefers the less coordinated phases.
This clearly shows that the change in the B-O bond length during an isotope exchange is the leading factor driving the production of 
the boron equilibrium isotope signatures at high $T$.

\section{Conclusions}

In this work we have presented a detailed analysis of boron isotope fractionation between boron-bearing crystalline solids
and aqueous fluids at high $T$ and $P$ conditions. In order to perform our investigation we have applied and extended a computationally efficient 
approach for the computation of 
isotope fractionation
factors for complex minerals and fluids at high temperatures and pressures presented by \citet{KJ11}.
As an extension to the \citet{BM47} {\it ``single atom approximation"} method we demonstrated 
that using the pseudofrequencies derived from the force constants acting on the fractionating 
element together with the full formula for computation of the reduced partition function ratios
results in significant improvement in the accuracy of the computed fractionation factors, which is essential 
when lower temperature materials and high vibrational frequency complexes are considered.

In order to understand the fractionation between B-bearing crystalline solids and aqueous fluids we performed a set of calculations
of $\beta$ factors for dravite, olenite, boromuscovite and aqueous solutions of $\rm H_3BO_3$ and $\rm H_4BO_4^-$.
In agreement with the experimental findings we show that the fractionation strongly correlates with coordination through the change in the B-O bond length.
The lower trigonal coordination $\rm BO_3$ arrangement results in higher $\rm ^{11}B/^{10}B$ (by $\rm \sim 8\,\permil $ at $T=\rm1000\,K$)
than the tetrahedrally coordinated boron complexes, which exhibit $\rm \sim0.15\,\AA$ longer B-O bonds.
The computed fractionation between minerals and fluids of the same coordination are in good agreement 
with experiments. However, we predict larger isotope fractionation between boromuscovite and $\rm H_3BO_3$ fluid (by at least a few $\permil$)
than was measured {\it in situ} at high $P$ by \citet{WM05} and \citet{MW08}, but that is consistent with measurements on natural samples.
We note that the presence of $\rm B^{[4]}$ in high-$P$ fluid could reconcile the {\it in situ} experimental results with our prediction and other measurements. 
This is expected from the experiments of \citet{STH05}, but requires further experimental confirmation. If true,
this would open the possibility for using the isotope fractionation techniques as a tool to measure the speciation of boron in fluids and crystalline solids.
We have also demonstrated that with our computational approach we are able to correctly predict the pressure-induced isotope fractionation 
for compressed aqueous fluids, which indicates the ability of {\it ab initio} methods to predict the isotopic signatures of highly compressed materials,
even those that are difficult to investigate experimentally.

Our study confirms that {\it ab initio} computer simulations are a useful tool not only for prediction but also understanding 
the equilibrium stable isotope fractionation processes between various phases, including aqueous solutions, at high pressures and temperatures.
They can nicely complement experimental efforts, provide unique insight into the isotope fractionation process on the atomic scale
and deliver data for conditions that are inaccessible by the current experimental techniques. 

\section*{Acknowledgements}
The authors wish to acknowledge financial support in the framework of DFG 
project no. JA 1469/4-1. Part of the calculations were performed on the IBM 
BlueGene/P JUGENE of the John von Neumann Institute for Computing (NIC).
We are also grateful the associate editor Edwin A. Schauble and anonymous referees for constructive comments that helped
improving the manuscript.





\bibliographystyle{model1-num-names}
\bibliography{<your-bib-database>}



\end{document}